\documentclass{article}

\usepackage{arxiv}

\usepackage[utf8]{inputenc} 
\usepackage[T1]{fontenc}    
\usepackage{hyperref}       
\usepackage{url}            
\usepackage{booktabs}       
\usepackage{amsfonts}       
\usepackage{nicefrac}       
\usepackage{microtype}      
\usepackage{lipsum}
\usepackage{graphicx}
\usepackage{natbib}
\usepackage{breqn}
\usepackage{amssymb}
\usepackage{amsfonts}
\usepackage{subcaption}
\usepackage{bbm}
\usepackage{bm}
\usepackage{algorithm, algpseudocode,enumitem}
\usepackage{xcolor}
\graphicspath{ {./images/} }

\title{Joint modeling of longitudinal HRQoL data accounting for the risk of competing dropouts}

\author{
 Hortense Doms \\
  LIDAM/ISBA, UCLouvain\\Louvain-la-Neuve, Belgium\\
  \texttt{hortense.doms@uclouvain.be} \\
   \And
 Catherine Legrand \\
  LIDAM/ISBA, UCLouvain\\Louvain-la-Neuve, Belgium\\
  \And
 Philippe Lambert \\
  LIDAM/ISBA, UCLouvain, Louvain-la-Neuve, Belgium\\
  Institut de Math\'ematiques, Universit\'e de Li\`ege, Belgium
}

\begin{document}
\maketitle
\begin{abstract}
In cancer clinical trials, health-related quality of life (HRQoL) is an important endpoint, providing information about patients' well-being and daily functioning. However, missing data due to premature dropout can lead to biased estimates, especially when dropouts are informative. 
This paper introduces the \texttt{extJMIRT} approach, a novel tool that efficiently analyzes multiple longitudinal ordinal categorical data while addressing informative dropout. 
Within a joint modeling framework, this approach connects a latent variable, derived from HRQoL data, to cause-specific hazards of dropout. Unlike traditional joint models, which treat longitudinal data as a covariate in the survival submodel, our approach prioritizes the longitudinal data and incorporates the log baseline dropout risks as covariates in the latent process. This leads to a more accurate analysis of longitudinal data, accounting for potential effects of dropout risks. Through extensive simulation studies, we demonstrate that \texttt{extJMIRT} provides robust and unbiased parameter estimates and highlight the importance of accounting for informative dropout. We also apply this methodology to HRQoL data from patients with progressive glioblastoma, showcasing its practical utility.\\
\end{abstract}

\keywords{Bayesian joint models \and Informative dropout \and  Item response theory \and Quality of life}

\section{Introduction}
\label{s:intro}
In cancer clinical trials, overall survival (OS) is the standard endpoint used to demonstrate the clinical benefit of tested treatments. However, while OS provides essential information on treatment efficacy, it may not fully reflect the impact of therapies on patients' daily well-being. To address this limitation, other endpoints have emerged, with health-related quality of life (HRQoL) becoming a crucial one. HRQoL data are collected using self-administered questionnaires at various time points during patient care and follow-up, providing valuable information on patients' perspectives on their well-being and daily functioning. These data are particularly important for evaluating the overall benefits of new therapies.
\\

\noindent 
Quality of life can be divided into various scales, each measured through one or more items. These items are often measured on a Likert scale, which generates ordinal data. A common approach to analyzing these data is the scoring method \citep{Fayers2001}. Item responses are summed to obtain a total score for each scale and each subject at different time points. While this scoring method is widely used, it has significant limitations \citep{GorterFox2015}. First, the HRQoL score is often treated as a continuous variable and analyzed using linear mixed models (LMM) under the assumption of normality. However, the total score behaves more like an ordinal variable and may show asymmetry due to ceiling or floor effects, violating these assumptions. Furthermore, the gaps between adjacent response categories on Likert scales (e.g., "not at all," "a little," "quite a bit," "very much") are not necessarily uniform, but the scoring method assumes they are. Additionally, subjects with different item responses can end up with the same total score, failing to distinguish between their distinct response patterns, which results in a loss of important information.\\

\noindent Given these challenges, more appropriate methods are needed to account for the ordinal nature of HRQoL data. Item response theory (IRT) models offer an alternative by linking individuals' item responses (raw data) to a latent variable representing the underlying HRQoL component of interest \citep{VanDerLinden2016}. In longitudinal analyses, some studies have extended IRT models by regressing the latent variable on predictors and incorporating subject-specific random effects \citep{Verhagen2013, WangDynamic2022}.\\

\noindent Another important issue with HRQoL data is dropout, which is particularly frequent in cancer clinical trials. Patients may stop completing questionnaires before the end of the clinical trial due to various reasons, such as disease progression or death. When dropout is related to such factors, it is considered informative and should be taken into account in HRQoL analyses to avoid biased results. To address this, joint models (JMs) can be used, as they can model the association between dropout and HRQoL outcomes. Several comprehensive overviews of the joint modeling framework are available \citep{tsiatis2004joint,rizopoulos2012joint,papageorgiou2019overview,wu2012analysis}. \citet{Touraine2023} demonstrated the importance of using joint models for HRQoL data in the presence of informative dropout. Similarly, \citet{Cuer2023} highlighted the need to distinguish between different causes of dropout by using competing risk joint models. It is indeed crucial to distinguish between informative and non-informative dropouts. Lastly, IRT models have been applied within the joint modeling framework for longitudinal ordinal data. For example, \citet{Luo2014} assessed the impairment caused by Parkinson's disease using a JM with a longitudinal IRT model but did not account for competing dropout risks.\\

\noindent In this paper, we aim to develop a model that allows the longitudinal analysis of multiple ordinal categorical outcomes while accounting for different dropout risks. We propose a joint modeling framework that uses an IRT model for the multivariate ordinal longitudinal outcomes and a proportional cause-specific hazards model for the dropout risks. Additionally, we aim to improve the analysis of longitudinal data by incorporating survival information directly into the longitudinal submodel. While classical joint models focus on survival information and treat longitudinal data as a covariate in the survival submodel, our approach prioritizes the longitudinal components by incorporating log baseline dropout risks as covariates in the latent longitudinal process. This enables us to account for the potential effects of dropout risks when estimating the longitudinal latent trait.\\

\noindent The rest of the paper is structured as follows: In Section 2, we describe the formulation of our proposed joint model. Section 3 presents a Bayesian estimation procedure for the model. In Section 4, we discuss a simulation study to assess the statistical performance of the model and Section 5 illustrates our approach by analyzing responses to the QLQ-C30 QOL questionnaires completed by glioblastoma patients. Finally, Section 6 concludes the paper with a discussion. 

\section{Methodology}
\label{s:methodology}

\subsection{Longitudinal submodel}
\noindent We consider a cohort of $n$ patients followed up over time. For every individual $(i = 1, \dots,n)$, we observe $K$ longitudinal ordinal outcomes measuring the same construct of interest, and each outcome contains $L_k$ categories. The longitudinal values are denoted $y_{ijk}$ for outcome $k$ $(k = 1\dots, K)$ observed at time $t_{ij}$ $(j = 1, \dots, N_i)$. 

\subsubsection{Latent process}
\noindent We defined $\eta_{ij}$ as latent variable that represents an underlying trait or ability of subject $i$ at occasion $j$ that cannot be directly observed but can be assessed through the longitudinal responses provided to the different items. The trajectory of $\eta$ over time for patient $i$ is described using a linear mixed model as follows : 
\normalsize
	\begin{equation}
	\left\lbrace
	\begin{array}{l}
	\eta_{i}(t) =  \bm{x}_i^T(t) \, \bm{\beta} + \bm{z}_i^T(t) \, \bm{b}_i \\
	\bm{b}_i \sim \mathcal{N}(0,D) 
	\end{array}\right.
	\end{equation} 

 \noindent Here, $\bm{\beta}$ denotes the $p$-dimensional vector of fixed effects associated with a time-dependent design vector $\bm{x}_i(t)$, which typically includes an intercept, a time-dependent slope, and can adjust for other covariates. Meanwhile, $\bm{b}_i$ represents the $q$-dimensional vector of random effects linked to $\bm{z}_i(t)$. The fixed effects term describes the average trajectory of the latent process at the population level, while the random effects term captures individual-specific deviations from this trajectory. 

\subsubsection{Graded response model}
\noindent We now define the link between the latent process \( \eta \) and each longitudinal response. To achieve this, we use the Samejima's ordinal response model, also known as the graded response model (GRM) \citep{samejima1969estimation}. The model incorporates two key parameters: for each outcome, there is an associated discrimination parameter \( a_k \), and a threshold parameter \( \bm{d}_k = [d_{k,1}, \dots, d_{k,L_k-1}] \). \\

\noindent We define the cumulative probability for the $k$th ordinal outcome as
\normalsize
\begin{equation}   
		\text{logit}\lbrace \text{P}(y_{ijk} \geq l | \eta_{ij}) \rbrace = a_{k}\eta_{ij} + d_{kl}\; \;
	\Leftrightarrow \; \;  \text{P}(y_{ijk} \geq l | \eta_{ij}) =  \frac{1}{1+\exp(-(a_{k}\eta_{ij} + d_{kl}))}
        \end{equation}
        
\noindent with $\eta_{ij}=\eta_i(t_{ij})$ denoting the latent ability level of subject $i$ at time $t_{ij}$ and $l = 1, \dots, L_{k}$. Parameter $a_k > 0$ describes how well item $k$ can differentiate between patients at different trait levels and parameters $d_{kl}$ represents the point on the latent trait scale $\eta_{ij}$ at which the probability of endorsing a response greater than or equal to category $l$ is $50\%$. Also, the order constraint  $d_{k,1} >  \dots  > d_{kl} >\dots > d_{k,L_{k}-1}$ must be satisfied. The boundary of response probabilities are $ \text{P}(y_{ijk} \geq 1 | \eta_{ij}) =  1$ and $\text{P}(y_{ijk} \geq L_k+1 | \eta_{ij}) =  0 $ and the probability of being in a particular category is $\text{P}(y_{ijk} = l|  \eta_{ij}) = \text{P}(y_{ijk} \geq l |  \eta_{ij}) - \text{P}(y_{ijk} \geq l+1 |  \eta_{ij})$.

\subsection{Survival submodel}
As introduced in Section 1, HRQoL data are often affected by premature termination of questionnaire completion, known in this context as dropout. Since dropout can be either informative (disease related) or non-informative (non-disease related) depending on its cause, it must be accounted for in HRQoL analyses to avoid biased results. To address this, we use a competing risks model for the time-to-dropout, allowing us to distinguish between two types of dropout: dropouts unrelated $(p = 1)$ or related $(p = 2)$ to a clinical event such as death or disease progression. We assume the latter to be informative, in contrast to the former which is assumed to be uninformative. The event indicator $\delta_{ip}$ in $\bm{\delta}{i} = [\delta_{i1}, \delta_{i2}]$ takes value $1$ if the $i$-th individual drops out due to cause $p$ before administrative censoring, and $0$ otherwise \citep{Legrand2017}. \\

\noindent Let $T_i^*$ be the time at which the dropout occurred for the $i$-th individual, we observe $T_i = \min(T_i^*, C_i)$, where $C_i$ is the right administrative censoring time (end of study). We model the time-to-dropout using a proportional cause-specific hazards model \citep{putter2007tutorial}. For patient $i$, the cause-specific hazard of dropout due to cause $p$ is given by
\begin{equation}
 h_{ip}(t)  = h_{0p}(t) \exp \big ( \bm{\gamma}_{p}^T \bm{w}_i + \alpha_p m_p( t, \bm{b}_i) \Big ) ,  \,  \, \, \, t >0 
\end{equation}

\noindent where $h_{0p}(t)$ denotes the cause-specific baseline hazard  for cause $p$ ($p=1,2$) and $\bm{w}_i = [w_{i1}, \dots, w_{iG}]^T$ is the vector of baseline survival covariates for subject $i$ with cause-specific regression coefficients $\bm{\gamma}_{p}$. Parameters $\alpha_p$ quantify the association between the longitudinal latent process and the survival process. Different associations functions $m(t, \bm{b}_i)$ can be specified. The current-value and the random effects association structures are frequently used in the literature \citep{rizopoulos2012joint}. These are respectively defined by $m(t, \bm{b}_i) = \eta_i(t)$ and $m(t, \bm{b}_i) = \bm{b}_i$. \\

\noindent In the joint modeling framework, the baseline hazard should not be left unspecified; therefore we define the logarithm of the baseline hazard functions using B-splines \citep{Lambert2005, rizopoulos2016r} :  
\begin{equation}
\log \lbrace h_{0p}(t) \rbrace =  \sum^{U_p}_{u =1} \gamma_{h_{0p},u} B_{u}(t)
\end{equation}
where $\gamma_{h_{0p,u}}$ is the regression coefficient associated to the $u$th function of a large B-spline basis associated to equidistant knots on $(0, \max_i T_i)$.

\subsection{Accounting for dropout risks to enhance the latent process}
Finally, we extend the previous model to refine the modeling of the latent process by incorporating survival information into the longitudinal submodel. Specifically, we aim to account for the potential effects of the different baseline hazards associated with competing dropout risks on the longitudinal latent process. To achieve this, we define and include the following covariate 
\begin{equation*}
\bm{v}^T(t_{ij}) = [\log h_{01}(t_{ij}), \dots, \log h_{0P}(t_{ij})],
\end{equation*}
in the latent process model. This modification enables us to capture the influence of these baseline hazards on the trajectory of the latent variable \(\eta_{ij}\), and subsequently on the longitudinal responses. This part is crucial to ensure that our model adequately reflects the complexities of the relationship between longitudinal HRQoL data and dropout events. It is also important to note that, since we introduce the logarithm of the baseline hazard functions as covariates to describe the latent process, we have to use the random effects association structure to ensure identifiability in the model. This is crucial because using an alternative structure, such as \( m\{ t, b_i \} = \eta_i(t) \), would introduce circular dependencies that hinder parameter estimation.  The extended model is therefore defined as:
\begin{equation}
\label{eq1}
\left\lbrace
\begin{array}{l}
h_{ip}(t) = h_{0p}(t) \exp \big( \bm{\gamma}_{p}^T \bm{w}_i + \alpha_p \bm{b}_i\big), \\
\eta_{ij} = \bm{x}_i^T(t_{ij}) \, \bm{\beta} + \bm{z}_i^T(t_{ij})\, \bm{b}_i + \bm{v}^T(t_{ij}) \bm{\lambda}, \\
\text{logit}\lbrace \text{P}(y_{ijk} \geq l | \eta_{ij}) \rbrace = a_{k}\eta_{ij} + d_{kl} 
\end{array}\right.
\end{equation}
where the parameter vector 
$\bm{\lambda} = [\lambda_1, \dots, \lambda_P]^T$
characterizes the association between the baseline risk of dropout for the different causes and the latent variable of interest \(\eta_{ij}\). An estimated \(\lambda_p\) which is not significantly different from zero indicates that the baseline risk of dropout due to cause \(p\) has no impact on the latent variable \(\eta_{ij}\). Conversely, a significantly positive \(\lambda_p\) implies that a higher baseline risk of dropout for cause \(p\) is associated with an increased value of \(\eta_{ij}\). As introduced earlier, we consider two competing risks of dropout: one unrelated to the progression of the disease (\(p=1\)) and another related to death or disease progression (\(p=2\)). In this case, we define 
$
\bm{v}^T(t_{ij}) = [\log {h}_{01}(t_{ij}), \log {h}_{02}(t_{ij})].
$
Moreover, the covariate vectors $\bm{w}_i$ and $\bm{x}_i(t_{ij})$ may share some or all covariates.
\section{Bayesian estimation}
We apply a Bayesian approach where inference is based on the posterior of the model parameters. In particular, we estimate the parameteres of the extended joint model (\ref{eq1}) using Markov Chain Monte Carlo (MCMC) methods. 
\subsection{Likelihood function} 
In the framework of joint models for longitudinal and survival data, the likelihood is derived under the common assumption that the longitudinal and survival processes are independent conditionally on the random effects $\bm{b}_i$. Moreover, the longitudinal measurements $\bm{y}_i$ of each individual are assumed independent given the random effects. \\

\noindent The expression of the marginal likelihood contribution for the $i^{th}$ subject is written as
\begin{equation}
\label{eq2}
 L_i(\bm{\theta}) = \int p(T_i, \delta_i |\bm{b}_i; \bm{\theta}_s, \bm{\beta} )  \big[\prod_{k=1}^{K}\prod_{j=1}^{N_i} p  \bigl ( y_{ijk} |\,\bm{b}_i ; \bm{\theta}_y \bigr ) \big] p (\bm{b}_i ; \bm{\theta}_b) \; db_i\\
\end{equation}
where $\bm{\theta} = [\bm{\theta}_s^T, \bm{\theta}_y^T, \bm{\theta}_b^T]^T$ contains $\bm{\theta}_s$ the parameters for the survival outcomes, $\bm{\theta}_y$ the parameters for the longitudinal outcome and  $\bm{\theta}_b$ the parameters of the random-effect covariance matrix. Function $p(.)$ generically denotes the corresponding probability or density function. The marginal log-likelihood function formulated for all subjects is obtained by $l(\bm{\theta}) = \sum_{i=1}^{n} \log  L_i(\bm{\theta})$.

\noindent The extended expression for the marginal likelihood contribution of the $i$th subject in the survival submodel is given by \citep{putter2007tutorial}
\begin{equation}
\label{eq3}
\begin{split}
p(T_i, \bm{\delta}_{i} |\bm{b}_i ; \bm{\theta}_s, \bm{\beta}) &= \prod_{p=1}^{P} \big[ h_{ip}(T_i | \bm{b}_i ; \bm{\theta}_s, \bm{\beta}) \big] ^{\delta_{ip}} \, S_{ip}(T_i |\bm{b}_i ; \bm{\theta}_s, \bm{\beta}) \\
&= \prod_{p=1}^{P} \Big[ h_{0p}(T_i) \exp \{ \bm{\gamma}_p^T \bm{w}_i  + \alpha_p m_p( T_i, b_i)\} \Big]^{\delta_{ip}}\\
& \; \;  \; \times \exp \Big(-\int_0^{T_i} h_{0p}(s) \exp \{ \bm{\gamma}_p^T \bm{w}_i + \alpha_p m_p( s, b_i)\} \, ds \Big)  \\
&= \prod_{p=1}^{P} \exp\Big[ \sum^{U}_{u =1} \gamma_{h_{0p},u} B_{u}(T_i) + \bm{\gamma}_p^T \bm{w}_i + \alpha_p m_p( T_i, b_i) \Big]^{\delta_{ip}}\\
& \; \;  \;   \times \exp \Big[ - \exp \Big\{ \bm{\gamma}_p^T \bm{w}_i \Big\} \int_0^{T_i} \exp \Big\{ \sum^{U}_{u =1} \gamma_{h_{0p},u} B_{u}(s)   +\alpha_p m_p( s, b_i)\Big\} \, ds \Big] \\ 
\end{split}
\end{equation}

\noindent The probability contribution in Eq. (\ref{eq2}) for the observed longitudinal value of the $k$th item response can be expressed as follows:
\begin{equation}
\label{eq4}
\begin{split}
p  \bigl ( y_{ijk} |\,\bm{b}_i ; \bm{\theta}_y \bigr )  &= \prod^{L_k}_{l=1} p(y_{ijk} = l)^{\mathbb{1}_{y_{ijk} = l}}\\
& = \prod^{L_k}_{l=1} \Big[ p(y_{ijk} \geq l |  \eta_{ij}) - p(y_{ijk} \geq l+1 |  \eta_{ij}) \Big]^{\mathbb{1}_{y_{ijk} = l}}\\
& = \Big[ 1 - \text{expit}(a_{k}\eta_{ij} + d_{k,1} )\Big]^{\mathbb{1}_{y_{ijk} = 1}} \\
& \; \times \prod^{L_k - 1}_{l=2} \Big[ \text{expit}(a_{k}\eta_{ij} + d_{k,l} ) - \text{expit}(a_{k}\eta_{ij} + d_{k,l+1} ) \Big]^{\mathbb{1}_{y_{ijk} = l}} \\
& \; \times \Big[ \text{expit}(a_{k}\eta_{ij} + d_{k,L_k})\Big]^{\mathbb{1}_{y_{ijk} = L_k}} 
\end{split}
\end{equation}

\noindent where $\text{expit}(x) = \frac{1}{1 + e^{-x}}$. \\

\noindent The normality assumption for the distribution of random effects implies that
\begin{equation}
\label{eq5}
p(\bm{b}_i ; \bm{\theta}_b) = (2\pi)^{-q/2} \det(\bm{D})^{-1/2} \exp(-\bm{b}_i^t \: \bm{D}^{-1} \bm{b}_i /2) 
\end{equation}
 where $q$ denotes the dimension of the random effect vector. \vspace{0.3cm}

\noindent A numerical method must be used for the evaluation of the integral in (\ref{eq2}) because it has no closed-form solution. We approximate it using the 15-point Gauss-Kronrod quadrature formula \citep{rizopoulos2016r}.

\subsection{Priors and posterior distributions}
\noindent Bayesian estimation of model parameters is based on the posterior distribution. The expression for the posterior distribution of the model parameters is derived under the assumptions defined in Section 3.1. Using Bayes theorem, we obtain
\begin{equation}
\label{eq6}
p(\bm{\theta}, \bm{b} \,|\,\bm{T}, \bm{\delta}, \bm{y}) \propto \prod_{i=1}^n  \prod_{j=1}^{N_i} \prod_{k=1}^{K} p(T_i, \bm{\delta}_{i} \, | \, \bm{b}_i; \bm{\theta}_s) \;  p(y_{ijk} \, | \bm{b}_i\, ; \bm{\theta}_y) \;  p(\bm{b}_i ; \bm{\theta}_b)  \:p(\bm{\theta})
\end{equation}
where  $\bm{\theta}$ denotes the full parameter vector with joint prior  $p(\bm{\theta})$.  The expressions for $p(T_i, \bm{\delta}_{i} \, | \,\bm{b}_i; \bm{\theta})$, $ p(y_{ijk} \, | \bm{b}_i\, ; \bm{\theta})$ and $p(\bm{b}_i ; \bm{\theta}_b) $ can be found in (\ref{eq3}), (\ref{eq4}) and (\ref{eq5}), respectively. \vspace{0.3cm}

\noindent We use standard non-informative prior distribution for $\bm{\theta}$ \citep{papageorgiou2019overview,Alsefri}. In particular, independent univariate diffuse normal priors are used for $\bm{\beta}$ (the vector of fixed effects in the longitudinal submodel), for $\bm{\gamma}$ (the vector of regression coefficients in the survival submodel) and for $\bm{\alpha}$ (the vector of association parameters). We let all discrimination parameters $\bm{a}$ have Uniform[0,5] as prior distribution to ensure non-negativity. Several strategies can be employed to impose order constraints on the thresholds $\bm{d}$. Here, we use a uniform prior distribution for each threshold, with $d_{k,l-1}$ serving as the upper boundary for the prior of $d_{k,l}$. For the variance-covariance matrix of the random effects $\bm{D}$, we take a Wishart prior. The flexibility of the log baseline hazards expressed as a linear combination of a large number of B-splines can be counterbalanced using a Gaussian Markov field prior for $\gamma_{h_{0p}}$ \citep{eilers1996flexible,Lambert2005} :

\begin{equation}
p(\bm{\gamma_{h_{0p}}}| \tau_{p})  \propto \tau_{p}^{\rho(\bm{K_{p}})/2} \exp \big(- \frac{\tau_{p}}{2} \, \bm{\gamma_{h_{0p}}}^T \, \bm{K_{p}} \, \bm{\gamma_{h_{0p}}} \big) \;\; \text{  and  } \;\; \tau_{p} \sim \text{Gamma}(\mathfrak{a}_p,\mathfrak{b}_p) 
\end{equation}
where $\tau_{p}$ is a roughness penalty parameter and $\bm{K_{p}}  = \bm{\Delta}_r^T \bm{\Delta}_r $ is the penalty matrix of rank $\rho(\bm{K_{p}}) = U_p$ associated with the baseline hazard function $h_{0p}$, where $\bm{\Delta}_r$ denotes the $r$th difference penalty matrix (with $r = 2$, say). The amount of smoothness is controlled by $\tau_{p}$. It is generally recommended to set $\mathfrak{a}_p$ equals to 1 and $\mathfrak{b}_p$ equals to a small number (say 0.005) \citep{lang2004bayesian}. \\

\noindent A random sample from the joint posterior in (\ref{eq6}) is obtained using a Metropolis-within-Gibbs algorithm with Gibbs steps for the variance parameter and Metropolis steps for the other parameters. The detailed steps of the algorithm are outlined in Algorithm~\ref{alg:one}.
\begin{algorithm}[ht!]
\caption{MCMC estimation of joint model using extJMIRT approach}
\label{alg:one}
\begin{algorithmic}
\State  \textbf{Input:} Longitudinal data \((y_{ijk}, t_{ij})\) for \(i = 1, \dots, n\) individuals, survival data \((T_i, \bm{\delta}_i)\), priors distributions for all parameters.\vspace{0.4em}
\State \textbf{Initialize:} Set initial values $\bm{\beta}^{0}, \bm{\lambda}^{0}, \bm{b}_i^{0}, \bm{a}^{0}, \bm{d}^{0}, \bm{D}^{0}, \bm{\gamma}_p^{0}, \bm{\gamma}_{h_0{_p}}^{0}, \tau_p^{0}, \bm{\alpha}_p^{0}$ for all $i$ and $p$.

\State Set number of MCMC iterations \(I\), adaptive phase \(A\), burn-in \(B\), and thinning \(T\). \\ Total number of iterations is $M=A+B+I$.\vspace{0.4em}
\For{\(m = 1, \dots, M\)}
    \State \textbf{Step 1: Update longitudinal submodel parameters.}
    \State \hspace{0.6em} \textbf{Update} \(\bm{\beta}^{m}\) using Metropolis-Hastings:
        \begin{itemize}[label=$\circ$,labelsep=0.2cm, left=1.5cm]
            \item A proposal \(\bm{\beta}^{*}\) is drawn from a normal proposal distribution.
            \item Compute acceptance ratio:
            \[
            R = \frac{\pi(\bm{\beta}^{*} \mid \text{other parameters})}{\pi(\bm{\beta^{m-1}} \mid \text{other parameters})}.
            \]
            \item Set \(\bm{\beta}^{m} = \bm{\beta}^{*}\) when $U \sim$ Uniform(0,1) $< R$. Otherwise, \(\bm{\beta}^{m} = 
 \boldsymbol{\beta}^{m-1}.\)

        \end{itemize}
    \State \hspace{0.6em} \textbf{Update} \(\bm{\lambda}^{m}\) using Metropolis-Hastings.
    \State \hspace{0.6em} \textbf{Update} \(\bm{a}^{m}\) and \(\boldsymbol{d}^{m}\) using Metropolis-Hastings.
    \State \hspace{0.6em} \textbf{Update} \(\bm{b}_i^{m}\) using Metropolis-Hastings.
    \State \hspace{0.6em} \textbf{Update} \(\bm{D}^{m}\) using Gibbs steps (see Appendix).\vspace{0.5em}
    
    \State \textbf{Step 2: Update survival submodel parameters.}
    \For {\(p = 1, \dots, P\)}
    \State \hspace{0.6em} \textbf{Update} \(\boldsymbol{\gamma}_p^{m}\), \(\boldsymbol{\gamma}_{h_{0p}}^{m}\), \(\boldsymbol{\alpha}_p^{m}\) using Metropolis-Hastings.
        \State \hspace{0.6em} \textbf{Update} \(\tau_p^{m}\) using Gibbs steps (see Appendix).
   \hspace{1.6em} \EndFor
 \State \textbf{Store} \(\boldsymbol{\theta}^{m} = (\boldsymbol{\beta}^{m}, \boldsymbol{\lambda}^{m}, \boldsymbol{b}_i^{m}, \boldsymbol{a}^{m}, \boldsymbol{d}^{m}, \boldsymbol{D}^{m}, \boldsymbol{\gamma}_p^{m},  \boldsymbol{\gamma}_{h_{0p}}^{m}, \tau_p^{m}, \boldsymbol{\alpha}_p^{m})\) for all $i$.
\EndFor \vspace{0.4em}
\State \textbf{Output:} Posterior samples \(\{\boldsymbol{\theta}^{m}\}_{m=\text{seq}(A + B + 1, M, \text{by} = T)}\).
\end{algorithmic}
\end{algorithm}
The covariance matrices of the normal proposal distributions for the random walk Metropolis algorithm are carefully defined using preliminary estimates for each submodel separately. These proposal distributions are further tuned during an adaptive phase consisting of $A$ iterations. Afterward, a burn-in period of $B$ iterations is performed, followed by $I$ additional iterations. Finally, the chains are thinned according to a thinning interval of $T$. We end up with chains of size $I/T$ for each parameter.

\section{Simulation}
In this section, we present simulation studies with different settings designed to evaluate the performance of our approach, \texttt{extJMIRT}, which estimates an extended joint model using a Bayesian framework to analyze ordered categorical data while accounting for the effects of different types of dropouts. 

\subsection{Simulation design}

\noindent \textbf{Setting I:} We generate 500 datasets, each with a sample size of \(n = 500\) subjects and a maximum of 20 visits per subject. We consider three ordinal outcomes (\(K = 3\)), each with 4 categories ($L_k=4$ for all $k$). The maximum follow-up time is set to 20. Data is generated according to the following extended joint model:
\[
\left\lbrace
\begin{array}{l}
h_{i1}(t) = h_{01}(t) \exp \big( \gamma_1  w_i + \alpha_1 b_{0i}\big)\\
h_{i2}(t) = h_{02}(t) \exp \big( \gamma_2  w_i + \alpha_2 b_{0i} \big)\\
\eta_{ij} = \beta_1 t_{ij} +  \beta_2  w_i + \lambda_1  \log {h}_{01}(t_{ij}) + \lambda_2 \log{h}_{02}(t_{ij})  + b_{0i} \\
\text{logit}\lbrace \text{P}(y_{ijk} \geq l | \eta_{ij}) \rbrace = a_{k}\eta_{ij} + d_{kl}
\end{array}\right.
\]
where \(l = 1, \dots, 4\), \(w_i \sim \text{Bin}(1,0.5)\) and \(b_{01} \sim N(0,1.5^2)\). The true values of the other parameters are provided in Table \ref{table1} (left panel). Parameters of the graded response model are simulated by generating thresholds from a uniform distribution, which are then cumulatively summed and adjusted to ensure decreasing and appropriately spaced thresholds for each item. Discrimination parameters are drawn from a log-normal distribution to ensure their positivity. This process guarantees realistic ordinal responses (comparable to HRQoL data) for the GRM framework. We set \( a_1 = 1 \) and \( d_{1,1} = 0 \) to standardize the discrimination of the first item and establish a reference point for the threshold parameters, thereby facilitating stable and interpretable estimates \citep{wang2019joint}. Also, as mentioned in Section 2.3, we consider two competing risks of dropout, denoted by \(p = 1, 2\). The baseline hazard functions were specifically designed to reflect the different time windows in which the competing dropouts ((i.e. premature discontinuation of QOL completion) are most likely to occur in the context of the application of Section 5. Function \( h_{01}(t) \) was constructed to generate a higher incidence of dropout of type \(1\) during the latter stages of follow-up, while \( h_{02}(t) \) increases the risk of a dropout of type \(2\) earlier in the follow-up period. This setup was chosen to reflect the situation encountered in the application (see Section 5), where type 1 dropouts represent disease-unrelated dropouts, and type 2 dropouts represent disease-related dropouts. Both types of dropouts are equally represented. Each subject is censored at the end of the study, leading to a right-censoring rate of 10\%. Simulated survival times were generated for every subject using the approach described in \citet{crowther2013simulating}. The cumulative hazard is evaluated using numerical quadrature, and survival times are generated using an iterative algorithm that employs the numerical root-finding method of \citet{brent1973some}. \\

\noindent \textbf{Setting II:} In this simulation setting, we generated datasets as described in Setting I, with the exception that we changed the true values of \(\bm{\lambda}\). First, we set \(\lambda_1\) and \(\lambda_2\) to zero (Setting II.a). This allowed us to evaluate the model's ability to detect the absence of effects related to the baseline hazards \(h_{01}(t)\) and \(h_{02}(t)\). We want to assess whether the model could correctly identify the non-significance of these covariates while maintaining unbiased estimates for the other parameters. Additionally, we conducted a variant of this setting (Setting II.b), where only the effect associated with the risk of disease-unrelated dropout was non-significant (\(\lambda_1 = 0\)).\\

\noindent \textbf{Setting III:} In the third simulation setting, we compare our approach to a more standard classical joint model for longitudinal quality of life (QOL) data. Specifically, we examine the consequences of fitting a simplified model that does not account for the competing risk of dropout or the association between the latent trait and the logarithms of the baseline hazard functions. The data were generated using the same model as in Setting I, but we fit the simplified model (\texttt{simpleJMIRT}):
\[
\left\lbrace
\begin{array}{l}
h_{i}(t) = h_{0}(t) \exp \big( \gamma  w_i + \alpha b_{0i}\big)\\
\eta_{ij} = \beta_1 t_{ij} +  \beta_2  w_i + b_{0i} \\
\text{logit}\lbrace \text{P}(y_{ijk} \geq l | \eta_{ij}) \rbrace = a_{k}\eta_{ij} + d_{kl}\\
\end{array}\right.
\]
We want to assess the consequences of incorrectly specifying a model that ignores the competing risk structure and the contributions of \(\lambda_1  \log {h}_{01}(t)\) and \(\lambda_2 \log{h}_{02}(t)\) in the latent process.\\

\noindent In each setting, to ensure convergence, we run the model estimation with 10,000 iterations, with an adaptive phase of 1,500 iterations and a burn-in of 1,500. The chains are thinned by a factor of 10, retaining 1,000 iterations for each parameter. Geweke statistics \citep{geweke1992evaluating} are used as a diagnostic tool for assessing chain convergence. For each regression parameter, we evaluate bias, root mean squared error (RMSE), and the effective coverage of 95\% credible intervals (COV).

\subsection{Simulation results}

\noindent The performance measures obtained with Setting I are shown in Table \ref{table1} (left panel). For each regression parameter, we observe that the estimated biases are relatively low, while the coverage probabilities (COV) are close to the nominal levels. In particular, the good performances obtained for the discrimination and threshold parameters confirm that the model effectively differentiates items and their different response categories. Moreover, in the survival submodels, the parameters $\alpha_1$ and $\alpha_2$, as well as $\gamma_1$ and $\gamma_2$, were well estimated, indicating that the model can effectively capture the distinct effects associated with the two competing risks of dropout. The same conclusion can be applied to parameters $\lambda_1$ and $\lambda_2$. \\

\noindent  The results of Setting II.a can be found in Table ~\ref{table2}. We observe that the estimates of $\lambda_1$ and $\lambda_2$ are very close to zero, as expected, with minimal bias and no significant deviation from 0. Additionally, the performance in the estimation of the others longitudinal parameters \(\beta_1\) and \(\beta_2\) remains reliable. The results of Setting II.b are presented in Table \ref{tableA} in Appendix.\\

\noindent The results of Setting III are presented in Table ~\ref{table1} (right panel). These were obtained using the \texttt{simpleJMIRT} approach, which does not account for the competing risks of dropout and estimates only a single hazard function for the dropout time. The estimated value of \(\gamma\) is close to the average of the true values of \(\gamma_1\) and \(\gamma_2\), with a similar observation for the estimate of \(\alpha\). However, compared to Setting I, the performance of parameter estimates in the graded response model of the longitudinal model is noticeably poorer, particularly in terms of effective coverage, which fails to achieve the nominal 95\% level due to biases and model misspecification. For instance, the bias for \(d_{31}\) rises from 0.001 to 0.118, and the coverage declines from 92.2\% to 36.0\%. This suggests that the simplified model has difficulty providing reliable estimates, especially for threshold parameters, likely because it fails to account for the association between the latent trait and the risk of dropout.\\

\noindent Overall, the simulation study demonstrates the good performance of the \texttt{extJMIRT} approach in analyzing ordered categorical longitudinal data in the presence of different dropout types. The results confirm that the model accurately identifies the non-significance of irrelevant effects while maintaining robust performance in estimating other parameters. Furthermore, these simulations show the importance of accounting for the competing risk structure of dropouts and the potential association between the latent trait and the risks of informative dropout.

\begin{table} 
\small \sf\centering  
\caption{\small Simulation results for S=500 replications of sample size $n = 500$ in Setting I (left table) and Setting III (right table) with the \texttt{extJMIRT} and the \texttt{simpleJMIRT} approaches respectively: bias, root mean square error (RMSE) and effective coverage (COV) of 95\% credible intervals for the regression parameters. Est. gives the estimated values of survival parameters in the misspecified model.}
\label{table1}
  \centering  
   \hfill
\begin{minipage}{0.45\textwidth} 
\begin{tabular}{ccccc}
\toprule
 &&\multicolumn{3}{c}{Setting I} \\ 
\cline{3-5}  \vspace{-0.2cm} &&&&\\
 &True & Bias & RMSE & COV \\ 
\hline \vspace{-0.18cm} &&&&\\
$a_2$& \:\,0.851& \:\,0.004 & 0.059 & 0.914\\
$a_3$& \:\,1.237& \:\,0.004 & 0.080 & 0.930\\
$d_{12}$& -1.440& -0.005 & 0.068 & 0.944\\
$d_{13}$& -1.962& -0.006 & 0.086 & 0.940\\
$d_{21}$& \:\,1.011&  -0.002 & 0.072 & 0.938\\
$d_{22}$& \:\,0.466& -0.006 & 0.070 & 0.938\\
$d_{23}$& -0.440& -0.006 & 0.071 & 0.944\\
$d_{31}$& \:\,1.043& \:\,0.001 & 0.097 & 0.922\\
$d_{32}$& \:\,0.214& -0.003 & 0.094 & 0.922\\
$d_{33}$& -0.621& -0.005 & 0.094 & 0.928\\
\hline  \vspace{-0.2cm}  &&&&\\
$\beta_1$& \:\,0.150& \:\,0.006 & 0.014 & 0.954\\ 
$\beta_2$& \:\,0.400& -0.003 & 0.158 & 0.942\\ 
$\lambda_1$& -0.250& \:\,0.010 & 0.074 & 0.972\\ 
$\lambda_2$& \:\,0.100& -0.008 & 0.080 & 0.974\\ 
\hline  \vspace{-0.2cm}  &&&&\\
$\gamma_1$& -1.000& -0.006 & 0.159 & 0.932\\ 
$\gamma_2$& -0.750& -0.009 & 0.140 & 0.934\\  
$\alpha_1$& -0.250& \:\,0.005 & 0.052 & 0.960\\  
$\alpha_2$& \:\,0.250& \:\,0.003 & 0.054 & 0.932\\  
\midrule
\end{tabular}
\end{minipage}
\hfill
\begin{minipage}{0.45\textwidth} 
 \begin{tabular}{ccccc}
\toprule
 &&\multicolumn{3}{c}{Setting III} \\ 
\cline{3-5}  \vspace{-0.2cm} &&&&\\
 &True & Bias & RMSE & COV \\ 
\hline \vspace{-0.18cm} &&&&\\
$a_2$& \:\,0.851& -0.018 & 0.057 & 0.896\\
$a_3$& \:\,1.237& -0.029 & 0.081 & 0.896\\
$d_{12}$& -1.440& \:\,0.035 & 0.067 & 0.880\\
$d_{13}$& -1.962& \:\,0.038 & 0.077 & 0.876\\
$d_{21}$& \:\,1.011& \:\,0.079 & 0.120 & 0.462\\
$d_{22}$& \:\,0.466& \:\,0.079 & 0.115 & 0.622\\
$d_{23}$& -0.440& \:\,0.083 & 0.116 & 0.634\\
$d_{31}$& \:\,1.043& \:\,0.118 & 0.174 & 0.360\\
$d_{32}$& \:\,0.214& \:\,0.121 & 0.167& 0.498\\
$d_{33}$& -0.621& \:\,0.124 & 0.168 & 0.522\\
\hline  \vspace{-0.2cm}  &&&&\\
$\beta_1$& \:\,0.150& -0.006 & 0.013 & 0.890\\ 
$\beta_2$& \:\,0.400& \:\,0.011 & 0.158 & 0.942\\ 
\hline  \vspace{-0.2cm}  &&&&\\
& Est.&  &   &   \\ 
\hline  \vspace{-0.2cm}  &&&&\\
$\gamma$&   -0.866 & / & / & /\\  
$\alpha$&    \:\,0.010 & / & / & /\\  
\midrule
\end{tabular}
\end{minipage}
\bigskip
\end{table}

\begin{table}
\small \sf\centering  
\caption{\small Simulation results for S=500 replications of sample size $n = 500$ in Setting II.a with the \texttt{extJMIRT} approach: bias, root mean square error (RMSE) and effective coverage (COV) of 95\% credible intervals for the regression parameters.}
\label{table2}
\begin{tabular}{ccccc}
\toprule
 &&\multicolumn{3}{c}{{Setting II.a}} \\
\cline{3-5}  \vspace{-0.2cm} &&&&\\
 &True & Bias & RMSE & COV \\ 
\hline \vspace{-0.18cm} &&&&\\
$a_2$& \:\,0.851& \:\,0.006 & 0.054 & 0.928\\
$a_3$& \:\,1.237& \:\,0.007 & 0.067 & 0.952\\
$d_{12}$& -1.440& -0.007 & 0.056 & 0.948\\
$d_{13}$& -1.962& -0.014 & 0.067 & 0.938\\
$d_{21}$& \:\,1.011& -0.007 & 0.068 & 0.918\\
$d_{22}$& \:\,0.466& -0.010 & 0.067 & 0.914\\
$d_{23}$& -0.440& -0.012 & 0.068 & 0.920\\
$d_{31}$& \:\,1.043& -0.012 & 0.081 & 0.924\\
$d_{32}$& \:\,0.215& -0.012 & 0.074 & 0.962\\
$d_{33}$& -0.621& -0.014 & 0.080 & 0.954\\
\hline  \vspace{-0.2cm}  &&&&\\
$\beta_1$& \:\,0.150& \:\,0.001 & 0.012 & 0.976\\ 
$\beta_2$& \:\,0.400& \:\,0.000 & 0.148 & 0.952\\ 
$\lambda_1$& \:\,0.000& -0.002 & 0.088 & 0.968\\ 
$\lambda_2$& \:\,0.000& \:\,0.000 & 0.092 & 0.970\\ 
\hline  \vspace{-0.2cm}  &&&&\\
$\gamma_1$& -1.000& -0.006 & 0.154 & 0.948\\ 
$\gamma_2$& -0.750& -0.005 & 0.136 & 0.938\\  
$\alpha_1$& -0.250& \:\,0.000 & 0.056 & 0.936\\  
$\alpha_2$& \:\,0.250& \:\,0.004 & 0.052 & 0.938\\  
\midrule
\end{tabular}
\end{table}
\normalsize 
\section{Application to glioblastoma data}

The dataset used in this study originates from a phase III clinical trial involving patients with first progression of glioblastoma. This trial was conducted by the European Organisation for Research and Treatment of Cancer (EORTC-26101 trial) to determine whether the combination of lomustine and bevacizumab (combination group) improved overall survival compared to lomustine alone (monotherapy group). The details of this trial can be found at \citet{clinic} (Identifier NCT01290939). While the primary endpoint of the trial is overall survival, these patients have a poor prognosis, making it crucial to evaluate the treatment's impact on their well-being.
To this end, quality of life data were collected longitudinally using the EORTC Quality of Life Questionnaire (QLQ-C30) \citep{Fayers2001} version 3, which assesses various dimensions of HRQoL. The QLQ-C30 is composed of 30 ordinal items that measure different aspects of quality of life, including a global health status/HRQoL scale, five functional scales (physical, role, emotional, social, and cognitive), three symptom scales (fatigue, nausea and vomiting, and pain), and six single items (dyspnea, insomnia, appetite loss, constipation, diarrhea, and financial difficulties). Patients self-reported their responses at multiple follow-up points: at baseline and every 12 weeks thereafter, allowing the monitoring of changes in HRQoL across different dimensions over time. \\

\noindent Trial results were already published \citep{Wick} and did not demonstrate a survival advantage of treatment with lomustine plus bevacizumab over lomustine alone in patients with progressive glioblastoma, despite a prolonged progression-free survival (PFS). Additionally, the health-related quality of life was analyzed using the sum-score approach, but no significant impact of treatment was found.\\

\noindent To demonstrate our methodology, we focus on the physical functioning scale, which consists of five items. They are presented in Appendix. Each item has a response range from 1 ("Not at all") to 4 ("Very much"), where higher responses indicate worse physical functioning. As described in previous sections, accurately estimating the latent process requires accounting for patient dropout and distinguishing between dropouts unrelated to the disease (cause 1) and those related to death or disease progression (cause 2). Following \citet{Cuer2023}, dropout was classified as cause 2 if death or progression occurred between the last questionnaire completed and the subsequent planned visit; otherwise, it was classified as cause 1. \\

\noindent The dataset analyzed includes $n = 423$ patients enrolled across multiple institutions in Europe. 
 Of these, $278$ were randomized to receive monotherapy treatment, while the rest were randomized to receive a combination of lomustine and bevacizumab. Follow-up times for the longitudinal HRQoL data ranged from 8 to 674 days. The median overall survival was 283 days, and the median progression-free survival was 95 days (see Figure A in the Appendix). Disease-unrelated dropout (cause 1) occurred in $31.2\%$ of patients, while $32.4\%$ experienced disease-related dropout (cause 2). The remaining patients, who do not drop out, were right-censored at the end of the study period. The cumulative incidence functions are shown in Figure B in the Appendix. Each patient completed the corresponding HRQoL questionnaire between 1 and 9 times during their follow-up, for an average of 2.18. \\

\noindent We consider the following joint model in our analysis:
\[
\left\lbrace
\begin{array}{l}
h_{i}^{\texttt{unr}}(t) = h_{0}^{\texttt{unr}}(t) \exp \big( \gamma_{\texttt{Trt}}^{\texttt{unr}}  w_{\texttt{Trt}} + \alpha^{\texttt{unr}} b_{0i}\big)\\
h_{i}^{\texttt{rel}}(t) = h_{0}^{\texttt{rel}}(t) \exp \big( \gamma_{\texttt{Trt}}^{\texttt{rel}} w_{\texttt{Trt}} + \alpha^{\texttt{rel}} b_{0i} \big)\\
\eta_{ij} = \beta_1 t_{ij} +  \beta_2  w_{\texttt{Trt}} + \lambda_1  \log {h}_{0}^{\texttt{unr}}(t_{ij}) + \lambda_2 \log{h}_{0}^{\texttt{rel}}(t_{ij})  + b_{0i} \\
\text{logit}\lbrace p(y_{ijk} \geq l | \eta_{ij}) \rbrace = a_{k}\eta_{ij} + d_{kl}
\end{array}\right.
\]

\noindent where $\eta_{ij}$ represents the latent physical functioning of patient $i$ at time $j$, with $l = 1, \dots, 4$ indicating the response categories and $k = 1, \dots, 5$ representing the five items of the physical functioning scale. Larger values of $\eta_{ij}$ correspond to worse physical functioning, while lower values indicate better functioning. \\

\noindent As presented in Section 3, we used Bayesian methods with Markov Chain Monte Carlo techniques to estimate this joint model. Trace plots revealed no discernible trends in the MCMC chains, suggesting adequate convergence. Furthermore, the Gelman-Rubin diagnostic confirmed that the potential scale reduction factor ($R$) for all parameters remained below 1.1, further validating convergence. The \texttt{extJMIRT} approach takes 4.23 minutes on a single core of a 2.6 GHz Intel Core i5-1145G7 processor, benefiting from a combination of R and C++ code for enhanced speed. The code is available upon request. The estimated parameters are presented in Table \ref{table3} for the survival and latent process parameters and in Table \ref{tableB} in the Appendix for GRM parameters.\\

\noindent The coefficients \(\beta_1\) and \(\beta_2\) represent the effects of time and treatment, respectively, on the latent trait \(\eta_{ij}\), which reflects physical functioning. The significant positive estimate of \(\beta_1\) suggests that, over the course of up to 674 days of follow-up, physical functioning tends to deteriorate. Additionally, the combined treatment does not appear to have a significant effect on \(\eta_{ij}\), suggesting that it may not impact the physical ability of patients. A similar result was observed in the original presentation of the study results \citep{Wick}. Parameters \(\lambda_1\) and \(\lambda_2\) capture the influence of the log-transformed baseline hazard functions for each type of dropout on the latent trait over time. The significance of both estimates underscores the importance of incorporating baseline risk effects in the specification of the latent trait, a process made feasible by our approach \texttt{extJMIRT}. Specifically, \(\lambda_2 = 0.122\) indicates that, at a given time point, a higher baseline risk of disease-related dropout is associated with worse physical functioning at the same time point, while a higher baseline risk of disease-unrelated dropout is associated with better physical functioning (\(\lambda_1 = -0.248\)). This latter finding may suggest that patients who experience improvements in their well-being may choose to discontinue follow-up visits because they no longer perceive a need for continued monitoring.\\

\noindent The estimated discrimination parameters (\(a_k\)) for the items range from 0.591 (\(a_4\)) to 1.349 (\(a_2\)). This indicates that \texttt{Item 2} has the highest ability to differentiate between patients with different levels of physical functioning, while \texttt{Item 4} is the least discriminative. The discrimination parameters for \texttt{Items 1, 3}, and \texttt{5} all take values around 1.00.  This pattern aligns well with the interpretation of the items' meanings: \texttt{Item 2}, which assesses "High Mobility", demands the highest physical effort, resulting in its strong discriminative power. \texttt{Items 1, 3}, and \texttt{5}, grouped as "Moderate Mobility" items, show moderate discrimination. Finally, \texttt{Item 4}, labeled as a "Low Mobility" item, is least discriminative as it simply reveals the need for rest during the day. Concerning the threshold parameters, we recall that \(d_{kl}\) represents the cut-off on the latent trait \(\eta_{ij}\) at which a patient is equally likely to choose response category \(l\) or higher, as compared to a lower category, for item \(k\). For example, for item 2, a patient with a latent trait value of \(\eta_{ij} = -2.607\) (i.e., \(d_{22}\)) has a 50\% probability of endorsing a response of 2 ("A Little") or higher. Higher \(d_{kl}\) values indicate that more severe functioning problems (higher \(\eta_{ij}\) values) are needed to endorse higher categories of response. \\

\noindent As an illustration, Figure 1 presents the estimated distribution of response probabilities for \texttt{Item 2} (\textit{Do you have any trouble taking a long walk?}), identified as the most discriminative item. Specifically, it illustrates the temporal evolution of the probabilities that an average patient ($b_{0i} = 0$) selects a specific response category $l \in \{1, \dots, 4\}$ at various time points $t \in \{100, 200, 300, 400, 500, 600\}$ (in days) after randomization, while accounting for the two types of dropout. As the treatment does not have a significant effect on $\eta_{ij}$, this covariate was taken out of the model. The black dots represent the estimated values of $\eta_{i}(t)$ for such a patient. At time $t = 100$, the patient is most likely to select the response category "A Little" for \texttt{Item 2}, with $\text{P}(Y_{100} = 2) = 53.5\%$. Over time, as the value of $\eta$ increases, the probability of selecting higher response categories rises, with "Quite a Bit" becoming the most likely category starting at $t=400$.\\

\begin{figure}
 \centerline{\includegraphics[width=0.7\textwidth]{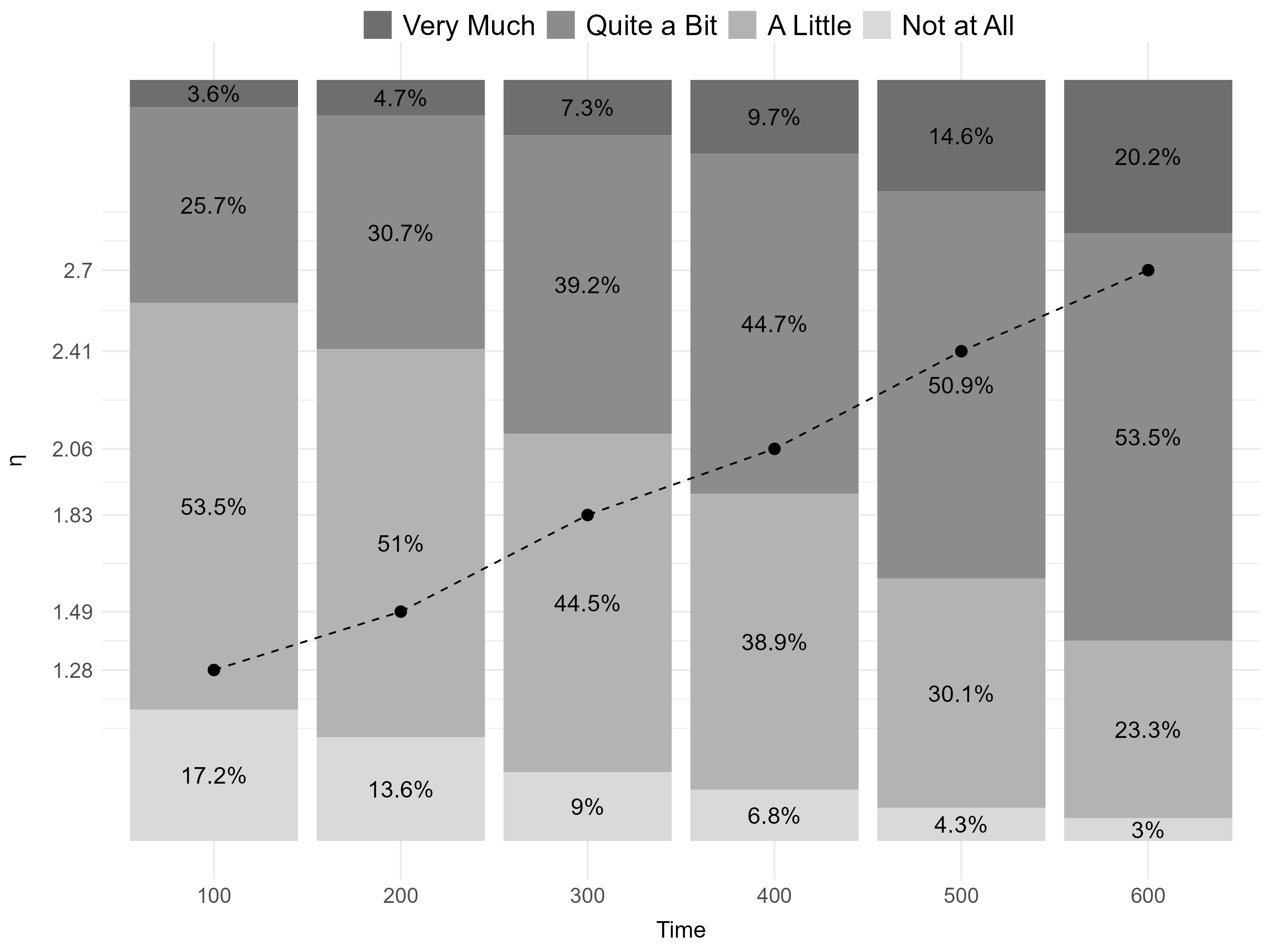}}
\caption{Distribution of response probabilities over time for \texttt{Item 2}. The stacked bars show the probabilities of selecting each response category at different times. Black dots represent the estimated values of $\eta_{i}(t)$ for an average patient $i$.} 
\label{fig1}
\end{figure}

\noindent Regarding the survival submodel parameters, the treatment demonstrates a significant negative effect ($\gamma_{\texttt{Trt}}^{\texttt{rel}} = -0.697$) on the risk of disease-related dropout, suggesting that patients receiving the combination treatment are less likely to withdraw due to death or progression of the disease compared to those receiving lomustine alone. This is consistent with previous findings \citep{Wick} showing that the combination treatment prolonged progression-free survival (PFS). In contrast, treatment has no significant effect on the risk of disease-unrelated dropout (see $\gamma_{\texttt{Trt}}^{\texttt{unr}}$ in Table \ref{table3}). This distinction shows the importance of understanding the reasons behind dropout. Moreover, the association between the individual-specific deviation from the latent trait and the risks of dropout is significant only in the disease-unrelated case ($\alpha^{\texttt{unr}} = 0.184$). \\

\noindent We also note the importance of distinguishing between the interpretations of \(\alpha_1\) and \(\lambda_1\) (and similarly for \(\alpha_2\) and \(\lambda_2\)). While both parameters assess the association between the risk of dropout and the latent trait, they approach it from different perspectives: one focuses on individual-specific effects, while the other examines the impact of the baseline dropout hazard on the latent trait. Together, they account for the complexity of the dropout process in longitudinal studies, especially in the context of assessing quality of life.

\begin{table}
\small \sf\centering 
\caption{  Estimation results of survival submodels and latent process parameters : estimates and 95\% credible intervals (CI) are presented.}
\label{table3}
\begin{tabular}{lcc}
\toprule
&\multicolumn{2}{c}{extJMIRT} \\
\hline
 & Est. & CI 95\% \\
\hline
$\beta_1$\;\;\;\,$\scriptsize{\texttt{(t}_{ij}\texttt{)}}$ &\:\,0.004&[\:\,0.002;\:\,0.005]\\ 
$\beta_2$\;\;\;\,$\scriptsize{\texttt{(Trt}_{\texttt{comb}}\texttt{)}}$ &\:\,0.284&[-0.271;\:\,0.820]\\ 
$\lambda_1$\;\;\;\,$\scriptsize{\texttt{(logh}_{0\texttt{unr}}\texttt{)}}$ &-0.248&[-0.399;-0.126]\\ 
$\lambda_2$\;\;\;\,$\scriptsize{\texttt{(logh}_{0\texttt{rel}}\texttt{)}}$ &\:\,0.122&[\:\,0.033;\:\,0.279]\\ 
\hline
$\gamma_{\texttt{Trt}}^{\texttt{unr}}$\,\;$\scriptsize{\texttt{(Trt}_{\texttt{comb}}\texttt{)}}$  &-0.380&[-0.807; 0.118]\\ 
$\gamma_{\texttt{Trt}}^{\texttt{rel}}$\,\;$\scriptsize{\texttt{(Trt}_{\texttt{comb}}\texttt{)}}$ &-0.697&[-1.066;-0.317]\\ 
$\alpha^{\texttt{unr}}$\,\,$\scriptsize{\texttt{(Assoc}\texttt{)}}$  &\:\,0.184&[\:\,0.105; 0.272]\\ 
$\alpha^{\texttt{rel}}$\,\,$\scriptsize{\texttt{(Assoc}\texttt{)}}$ &\:\,0.019&[-0.070; 0.109]\\ 
\hline
\end{tabular}
\end{table}
\section{Discussion}

\noindent The primary aim of this study was to develop the \texttt{extJMIRT} approach to efficiently analyze multiple longitudinal ordinal categorical data while accounting for potential informative dropout. This methodology is framed within a joint modelling approach that connects a latent variable, derived from multiple longitudinal ordinal outcomes in an IRT model, to cause-specific hazard of dropout. The latent variable is regressed on random effects and predictors, including survival information. Unlike traditional joint models, which treat longitudinal data as a covariate in the survival submodel, our approach prioritizes the longitudinal components by incorporating the log baseline risks of both competing dropout events as covariates in the latent process model. The \texttt{extJMIRT} approach thus provides a novel and robust tool for analyzing multiple ordinal categorical data, particularly in health-related quality of life assessments in the presence of (potentially informative) dropout.\\

\noindent The simulation studies showed that the proposed inference strategy effectively estimates the model parameters in an \texttt{extJMIRT} model with minimal or no bias. In the first setting, we demonstrated the model's ability to differentiate between response categories and accurately estimate the distinct effects associated with the two competing risks of dropout. In Setting II, the model successfully detected the absence of dropout effects on longitudinal ordinal outcomes, while maintaining accurate estimation of other parameters. Setting III highlights the importance of accounting for patient dropout and distinguishing between its different causes to obtain accurate estimates of the latent process underlying the multivariate longitudinal ordinal responses. \\

\noindent In our application to glioblastoma data, the \texttt{extJMIRT} approach provided a comprehensive analysis of HRQoL data and successfully captured the impact of disease-related dropout on patients' physical functioning over time. Our findings revealed that a higher baseline risk of disease-related dropout was associated with worse physical functioning, whereas disease-unrelated dropout was linked to better functioning. This underscores the necessity of accounting for different dropout mechanisms when analyzing longitudinal quality of life data in clinical trials.\\

\noindent Studying the time and reasons for patient dropout is critical, as it provides valuable information that can influence the analysis and interpretation of longitudinal data. Ignoring this informative aspect or failing to distinguish between different causes of dropout may lead to biased estimates and an incomplete understanding of the true effects of covariates. Although some clinical trials have started to include the collection of dropout reasons in their protocols, this information is still often incomplete, and the causes of dropout are often inferred retrospectively. Systematically gathering reasons for incomplete HRQoL questionnaires would improve the attribution of dropout causes and enhance the application of the \texttt{extJMIRT} approach.\\

\noindent Within the framework of joint models, several extensions are possible. Although we specifically accounted for competing dropout events, this methodology can accommodate other competing types of events of interest. In this paper, we modeled the categorical ordinal data using a graded response model, but other models within the IRT framework could easily be employed, depending on the structure of the raw data \citep{VanDerLinden2016}. Also, while our application focused on HRQoL data, the \texttt{extJMIRT} approach is not limited to this context and can be applied to any longitudinal ordinal categorical data, including other types of patient-reported outcome measures (PROMs). Additionally, more complex, non-linear trajectories in the evolution of the latent trait could be captured using spline-based approaches. Finally, we assumed that the multiple items studied reflect a univariate latent variable. Exploring a multidimensional latent trait model can provide valuable insights into different aspects of multivariate ordinal outcomes \citep{wang2019joint}. However, adding multiple longitudinal latent traits to capture these dimensions would substantially increase the computational complexity.

\section*{Acknowledgements}

The authors acknowledge the support of the ARC project IMAL (grant 20/25-107) financed by the Wallonia-Brussels Federation and granted by the Acad\'emie universitaire Louvain. The authors thank the European Organization for Research and Treatment of Cancer for permission to use the data from EORTC-26101. The contents of this publication and methods used are solely the responsibility of the authors and do not necessarily represent the official views of the EORTC.

\section{Appendix}
\setcounter{table}{0}
\setcounter{figure}{0}
\renewcommand{\thefigure}{\Alph{figure}}
\renewcommand{\thetable}{\Alph{table}}
\subsection*{Conditionals distributions used in the Gibbs sampling}
 \[
    \tau_p|\bm{\gamma_{h_{0p}}} \sim \text{Gamma} \left(\mathfrak{a}_p+\frac{\rho(\bm{K}_p)}{2}, \mathfrak{b}_p+\frac{\bm{\gamma_{h_{0p}}}^T \, \bm{K_{p}} \, \bm{\gamma_{h_{0p}}}}{2} \right),\\[-1.5cm]
    \]
    \[
    \boldsymbol{D|b_{i}} \sim \text{Inverse-Wishart}\left(q+n, q\times D^{0} + \sum_{i=1}^n \boldsymbol{b}_i \boldsymbol{b}_i^{\top} \right)
    \]
    where \(q\) is the dimension of the random effects $\bm{b}_i$.

\subsection*{Physical functioning scale}
\noindent Item 1: \textit{Do you have any trouble doing strenuous activities, like carrying a heavy shopping bag or a suitcase?}\\
\noindent Item 2: \textit{Do you have any trouble taking a \underline{long} walk?}\\
\noindent Item 3: \textit{Do you have any trouble taking a \underline{short} walk outside of the house?}\\
\noindent Item 4: \textit{Do you need to stay in bed or a chair during the day?}\\
\noindent Item 5: \textit{Do you need help with eating, dressing, washing yourself or using the toilet?}\\[0.2cm]
\noindent Possible answers : 1 = "Not at All", \,2="A Little", \,3="Quite a Bit", \,4="Very Much"
    
\begin{table}[htb]
\small \sf\centering  
\caption{\label{tableA} \footnotesize Simulation results for S=500 replications of sample size $n = 500$ in Setting II.b with the \texttt{extJMIRT} approach: bias, root mean square error (RMSE) and effective coverage (COV) of 95\% credible intervals for the regression parameters.}

\begin{tabular}{ccccc}
\hline
 &&\multicolumn{3}{c}{Setting II.b} \\
\cline{3-5}  \vspace{-0.2cm} &&&&\\
 &True & Bias & RMSE & COV \\ 
\hline \vspace{-0.18cm} &&&&\\                   
$a_2$& \:\,0.851& \:\,0.010 & 0.056 & 0.922\\
$a_3$& \:\,1.237& \:\,0.005 & 0.071 & 0.940\\
$d_{12}$& -1.440& -0.009 & 0.057 & 0.948\\
$d_{13}$& -1.962& -0.014 & 0.069 & 0.942\\
$d_{21}$& \:\,1.011& -0.008 & 0.070 & 0.856\\
$d_{22}$& \:\,0.466& -0.010 & 0.069 & 0.906\\
$d_{23}$& -0.440& -0.011 & 0.071 & 0.916\\
$d_{31}$& \:\,1.043& -0.017 & 0.092 & 0.854\\
$d_{32}$& \:\,0.214& -0.016 & 0.088 & 0.908\\
$d_{33}$& -0.621& -0.016 & 0.087 & 0.924\\
\hline  \vspace{-0.2cm}  &&&&\\
$\beta_1$& \:\,0.150& \:\,0.001 & 0.012 & 0.970\\ 
$\beta_2$& \:\,0.400& -0.008 & 0.146 & 0.995\\ 
$\lambda_1$& \:\,0.000& \:\,0.004 & 0.051 & 0.979\\ 
$\lambda_2$& \:\,0.100& -0.006 & 0.063 & 0.978\\ 
\hline  \vspace{-0.2cm}  &&&&\\
$\gamma_1$& -1.000& -0.005 & 0.156 & 0.950\\ 
$\gamma_2$& -0.750& -0.007 & 0.139 & 0.942\\  
$\alpha_1$& -0.250& -0.003 & 0.054 & 0.944\\  
$\alpha_2$& \:\,0.250& \:\,0.008 & 0.054 & 0.936\\  
\hline
\end{tabular}
\end{table} 
\begin{table}[H]
\small \sf\centering 
\caption{Estimation results of GRM submodel parameters : estimates and 95\% credible intervals (CI) are presented.}
\label{tableB} 
\begin{tabular}{llcc}
\hline
&&\multicolumn{2}{c}{extJMIRT} \\
\hline
& & Est. & CI 95\% \\
 \hline
&$a_1$  & 1.000 & / \\ 
&$a_2$  & 1.349 & [1.157;\:\,1.547] \\ 
&$a_3$  & 1.082 & [0.929;\:\,1.224] \\ 
&$a_4$  & 0.591 & [0.501;\:\,0.681] \\ 
&$a_5$  & 1.061 & [0.893;\:\,1.208] \\ 
\texttt{Item 1}&$d_{11}$  & 0.000 & / \\ 
&$d_{12}$  & -2.478 & [-2.746;\:\,-2.226] \\ 
&$d_{13}$  & -4.468 & [-4.861;\:\,-4.079] \\ 
\texttt{Item 2}&$d_{21}$  & -0.158 & [-0.464;\:\,-0.133] \\ 
&$d_{22}$  & -2.607 & [-3.012;\:\,-2.217] \\ 
&$d_{23}$  & -5.013 & [-5.605;\:\,-4.469] \\ 
\texttt{Item 3}&$d_{31}$  & -2.601 & [-2.936;\:\,-2.220] \\ 
&$d_{32}$  & -4.872 & [-5.295;\:\,-4.392] \\ 
&$d_{33}$  & -6.582 & [-6.973;\:\,-6.011] \\ 
\texttt{Item 4}&$d_{41}$  & -0.659 & [-0.861;\:\,-0.469] \\ 
&$d_{42}$  & -2.765 & [-3.044;\:\,-2.498] \\ 
&$d_{43}$  & -4.730 & [-5.255;\:\,-4.294] \\ 
\texttt{Item 5}&$d_{51}$  & -3.727 & [-4.118;\:\,-3.254] \\ 
&$d_{52}$  & -5.217 & [-5.676;\:\,-4.643] \\ 
&$d_{53}$  & -6.477 & [-6.965;\:\,-5.770] \\ 
\hline
\end{tabular}
\end{table}
\begin{figure}[H]
 \label{figureA}
    \centering
    \begin{subfigure}[t]{0.40\textwidth}
        \centering
        \includegraphics[width=\textwidth]{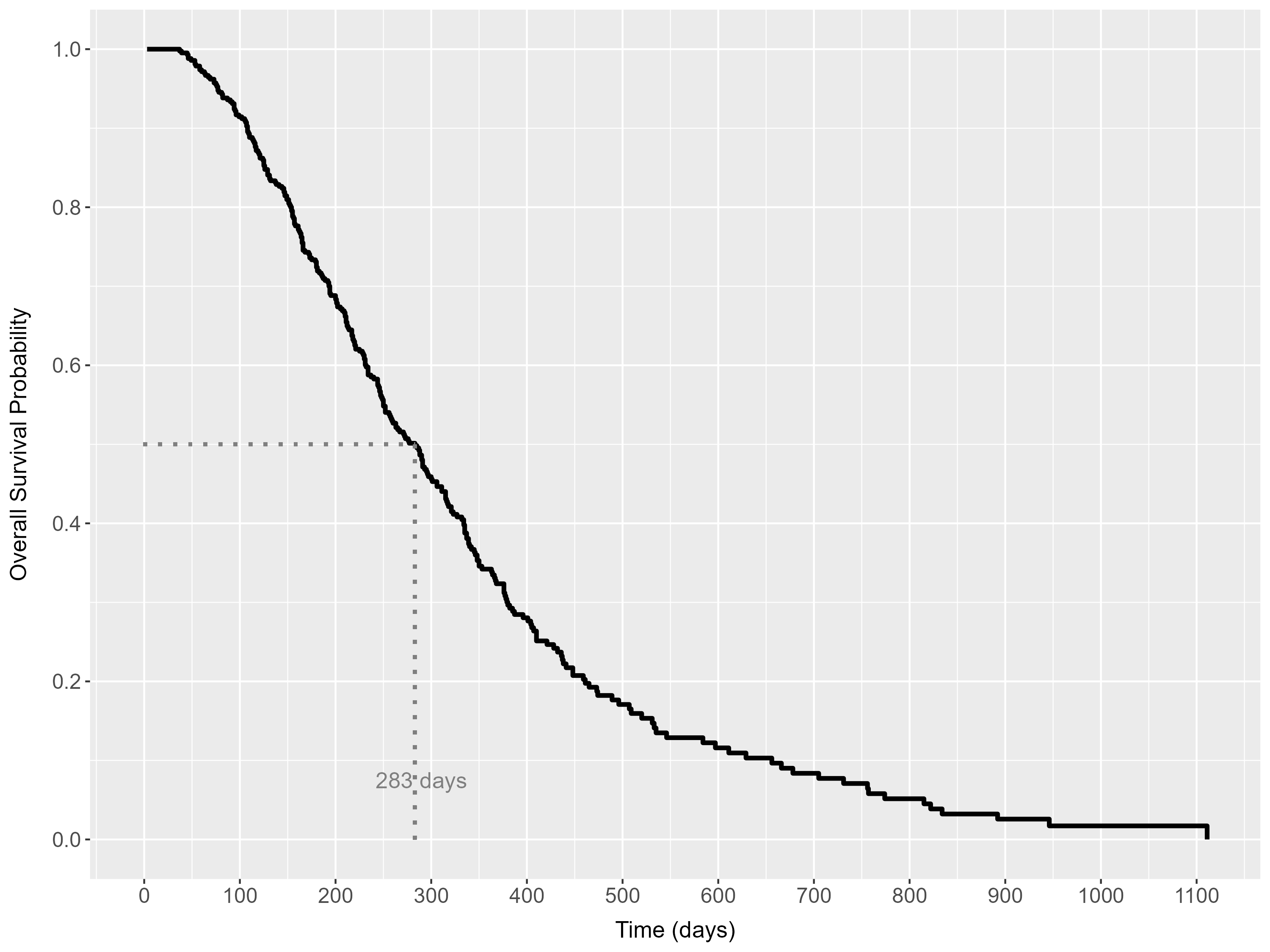}
    \end{subfigure}
    \hfill
    \begin{subfigure}[t]{0.40\textwidth}
        \centering
        \includegraphics[width=\textwidth]{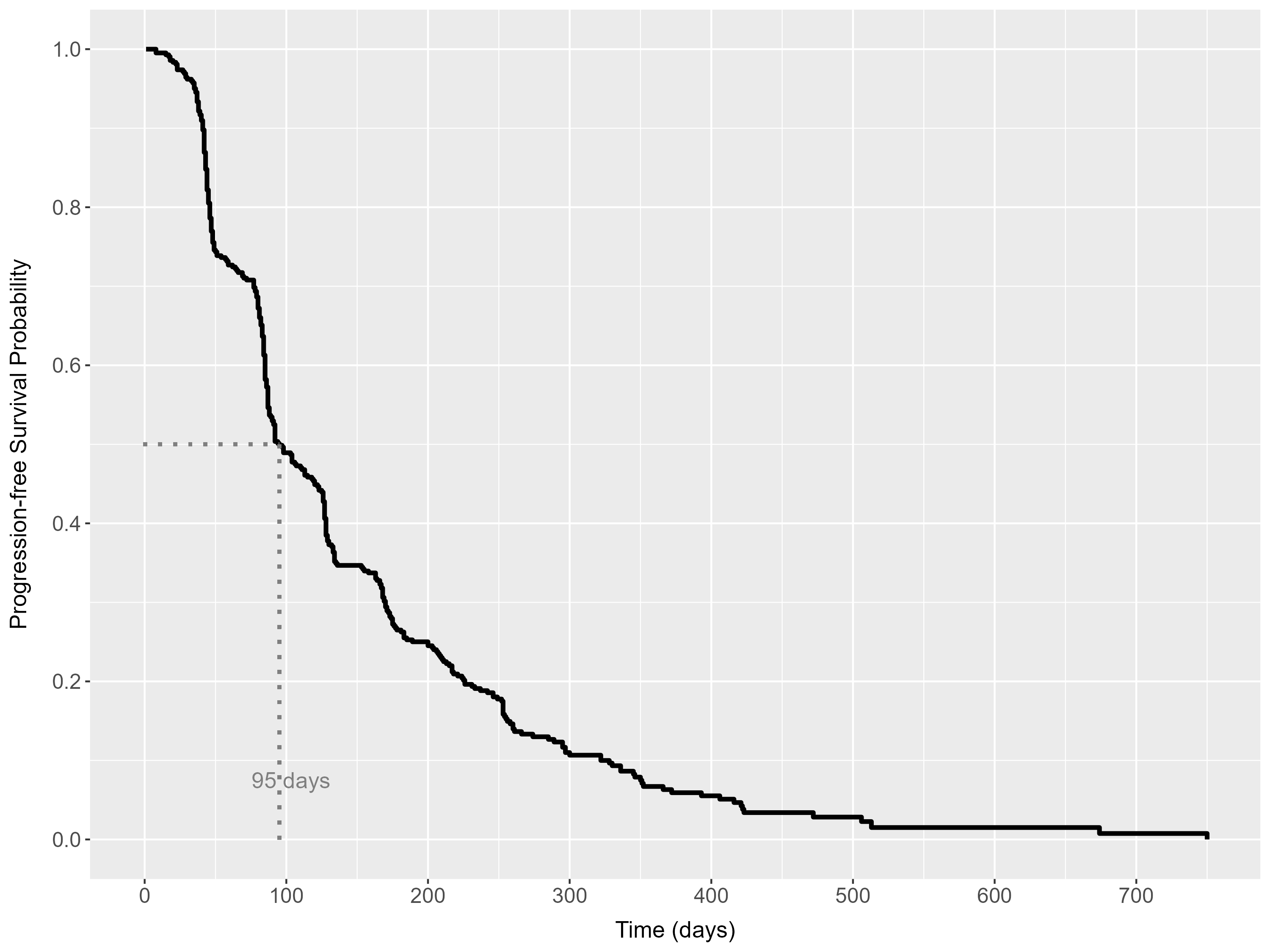}
    \end{subfigure}
    \caption{Overall Survival (left panel) and Progression-free Survival (right panel) in patients with first progression of glioblastoma}
    \label{fig:two_images}
\end{figure}
\begin{figure}[H]
        \centering
        \includegraphics[width=0.4\textwidth]{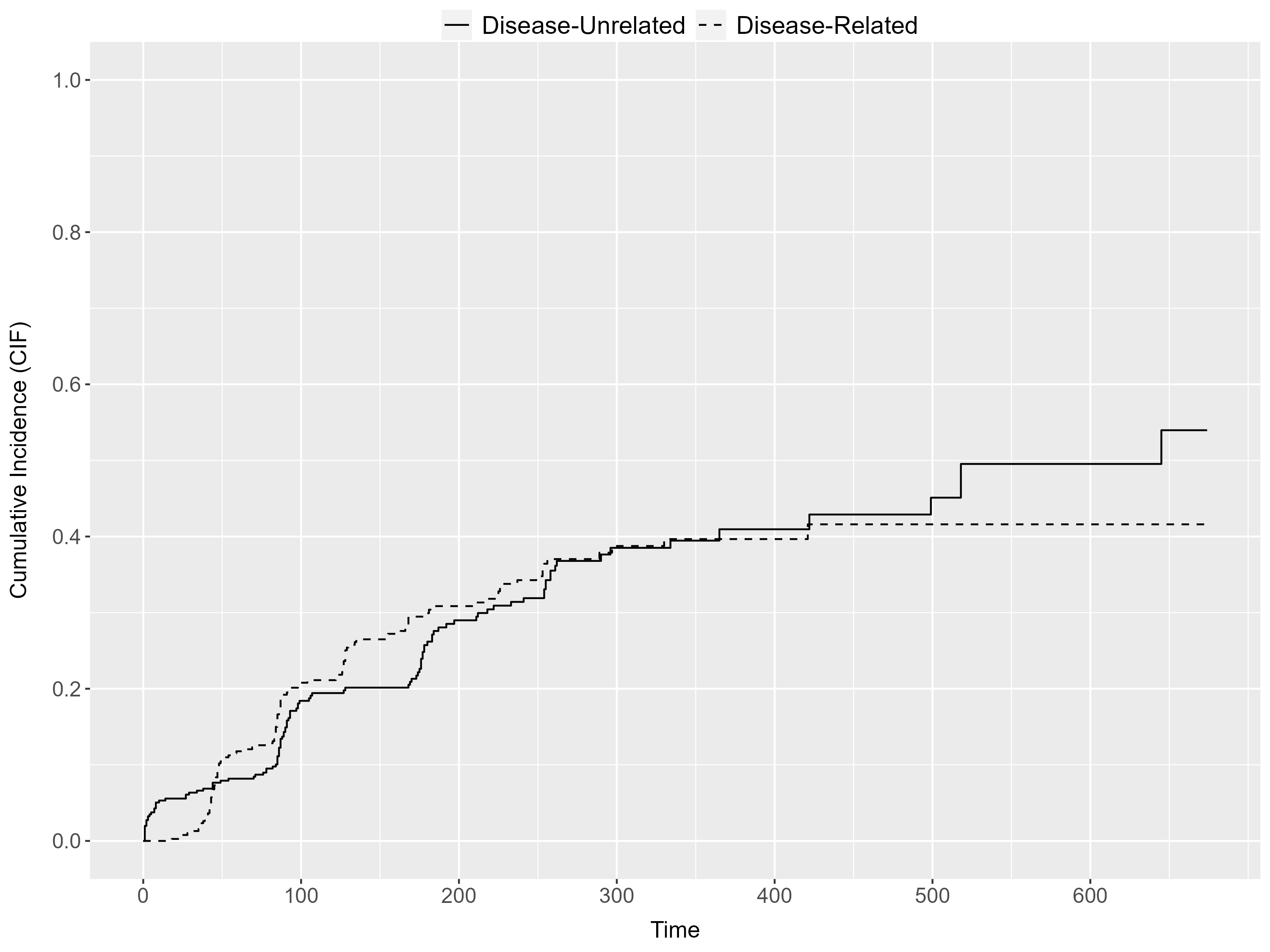}
        \label{figureB}
         \caption{Cumulative incidence functions for dropouts unrelated and related to disease progression or death.}
    \end{figure}

\end{document}